\pdfoutput=1
\documentclass[aps,prd,amsmath,floats,floatfix, twocolumn,
superscriptaddress,nofootinbib,showpacs,longbibliography]{revtex4-1}
\usepackage{float}
\usepackage[T1]{fontenc}
\usepackage[utf8]{inputenc}
\usepackage{lmodern}
\usepackage{verbatim}
\usepackage{multirow}
\usepackage{amsmath}
\usepackage[dvipsnames, usenames]{xcolor}
\definecolor{linkcolor}{rgb}{0.0,0.3,0.5}
\usepackage[hypertexnames=false, unicode, colorlinks=true, linkcolor=linkcolor,
citecolor=linkcolor, filecolor=linkcolor,urlcolor=linkcolor,
pdfusetitle]{hyperref}

\usepackage[all]{hypcap}
\usepackage{graphicx}
\usepackage{xspace}
\usepackage{amssymb}
\usepackage{microtype}
\usepackage[english]{babel}
\usepackage{blindtext}
\usepackage[normalem]{ulem} 
\usepackage{bm} 
\usepackage{mathrsfs}
\usepackage[nomessages]{fp}

\DeclareMathAlphabet{\mathpzc}{OT1}{pzc}{m}{it}

\colorlet{RED}{red}

\allowdisplaybreaks

\graphicspath{
	{plots/}
}
\begin{document}

\title{White dwarf mass-radius relation in theories beyond general relativity}
\newcommand{\UMassDMath}{\affiliation{Department of Mathematics,
		University of Massachusetts, Dartmouth, MA 02747, USA}}
\newcommand{\UMassDPhy}{\affiliation{Department of Physics,
		University of Massachusetts, Dartmouth, MA 02747, USA}}
\newcommand{\CSCVR}{\affiliation{Center for Scientific Computing and Visualization 	Research, University of Massachusetts, Dartmouth, MA 02747, USA}}
\newcommand{\IISER}{\affiliation{Department of Physical Sciences, Indian Institute of Science Education and Research Kolkata, Mohanpur - 741 246, WB, India}}    
\author{Khursid Alam}
\IISER
\author{Tousif Islam}
\UMassDPhy
\UMassDMath
\CSCVR
\hypersetup{pdfauthor={Alam et al.}}
\date{\today}

\begin{abstract}
We explore the internal structures of the white dwarfs in two different modified theories of gravity: (i) scalar-tensor-vector gravity and (ii) beyond Horndeski theories of $G_3$ type. The modification of the gravitational force inside the white dwarf results in the modification of the mass and radius of the white dwarf. We use observational data from various astrophysical probes including \textit{Gaia}  to test the validity of these two classes of modified theories of gravity. We update the constraints on the parameters controlling the deviation from general relativity (and Newtonian gravity in the weak field limit) as : $0.007 \le \alpha$ for the scalar-tensor-vector gravity and $-0.08 \le \gamma \le 0.007$ for the beyond Horndeski theories of $G_3$ type. Finally, we demonstrate the selection effect of the astrophysical data on the tests of the nature of gravity using white dwarf mass-radius relations specially in cases where the number of data-points are not many.
\end{abstract}

\maketitle
\section{Introduction} 
\label{Sec:Introduction}
General relativity (GR) is a highly successful theory because of its conceptual simplicity and geometrical interpretation, and various experimental tests have verified it \cite{Berti:2015itd}. But, there are some fundamental issues in its theoretical understanding. For example, GR fails to explain spacetime singularities in gravitational collapse \cite{Misner:1973prb}. At fundamental level, the quantum nature of all other forces are well understood except gravity, or more specifically GR.

When GR is used to describe the dynamics of the Universe, it is crucial to understand the content of the Universe. With the known baryonic content of the Universe, GR fails to clarify many observational phenomena like the rotation curves of the galaxies, mass profiles of the galaxy clusters, gravitational lensing effects, structure formation, and several cosmological data \cite{Joyce:2014kja}. The standard cosmological model, i.e. $\Lambda$CDM, introduced a cosmological constant (dark energy) and cold dark matter to explain cosmic acceleration, galaxy rotation curves and structure formation. But, we do not understand why the scale of the required cosmological constant is so small compared to its theoretically expected value and how gravity couples with the vacuum energy \cite{Martin:2012bt}. At the same time, there is no convincing non-gravitational evidence of cold dark matter. 

On the other hand, it is quite possible that the description of gravitational force is non-Einsteinian in nature \cite{Clifton:2011jh}. In this case, the role of dark matter and dark energy is mimicked by modified gravitational equations. Modified theories of gravity also serve as important test beds to analyse how well GR agrees with experiments. Motivated by these above reasons, many modified gravity theories have been proposed like the Modified Newtonian Dynamics theory (MOND) \cite{Milgrom:1983ca,Famaey:2011kh}, scalar-tensor-vector gravity theory (STVG) of Modified Gravity (MOG) \cite{Moffat:2005si}, beyond Horndeski theories of $G^{3}$ type \cite{Horndeski:1974wa}, Weyl Conformal gravity \cite{Mannheim:1988dj, Mannheim:2005bfa}, etc. These modified gravity theories succeed in explaining the dynamics of galaxies and globular clusters without invoking dark matter \cite{Islam:2019irh, Islam:2019szu, Islam:2019iua}. Modified gravity theories have successfully explained the observed rotation curves of a large selection of galaxies (\cite{Sanders:2002pf,Gentile:2010xt} for MOND; \cite{Mannheim:2010xw, Mannheim:1996rv, Mannheim:2010ti, OBrien:2011vks, Dutta:2018oaj,Islam:2018ymd} for Weyl gravity; \cite{Green:2019cqm} for MOG; for beyond Horndeski theories of $G^{3}$ type \cite{Koyama:2015oma}). All these modified gravity theories should also be verified for their predictions for some compact objects like binary pulsars, white dwarfs and neutron stars. 
Recently, white dwarfs have been used to test the astrophysical viability of various modified theories of gravity~\cite{Jain:2015edg,Banerjee:2017uwz,Kalita:2023hcl}.

In this paper, we want to specifically consider two modified gravity theories: MOG \cite{Moffat:2005si} and beyond Horndeski theories of $G^3$ type \cite{Zumalacarregui:2013pma, Gleyzes:2014dya, Gleyzes:2014qga}.  In MOG, the gravitational coupling constant $G$ is considered to be a scalar field whose numerical value usually exceeds Newton's constant ($G_{N}$). This theory can correctly explain galaxy rotation curves \cite{Brownstein:2005zz}, clusters dynamics \cite{Moffat:2013sja}, Bullet Cluster phenomena \cite{Brownstein:2007sr}, and also cosmological data \cite{Moffat:2007ju}, without considering the existence of dark matter and dark energy. Similarly, there exists a healthy extension of Horndeski theory via beyond Horndeski theory which can explain cosmic acceleration of the present universe and is a viable competitors of the $\Lambda$CDM model \cite{Kase:2014yya, Barreira:2014jha}. 
In this work, we use the equation of state of the white dwarf to solve two different modified Tolman-Oppenheimer-Volkoff (TOV) equations of the above mentioned alternative theories of gravity. We first calculate the mass-radius relations of the white dwarfs. These theoretical masses and radii of white dwarfs are then subsequently compared with the observed white dwarfs mass-radius data. Finally, we find the best-fitted parameters of the modified gravity theories by performing chi-square fitting on white dwarf data. 

This work is organized as follows: in Sec. \ref{White_dwarfs_in_beyond_GR_theories}, we briefly discuss the structure of white dwarfs in general relativity (GR), scalar-tensor-vector gravity theory or MOG theory and beyond Horndeski theories of types $G^{3}$. In Sec. \ref{Equation_of_state}, we discuss the equation of state of the white dwarfs while we comment on the observational mass-radius data of white dwarfs in Sec. \ref{sec:data}. Finally, in Sec. \ref{Result}, we provide an outline to calculate the mass-radius of the white dwarf and perform $\chi^2$ fitting to compare the observational mass-radius data with the theoretically predicted values. We draw our conclusions and discuss possible caveats in our analysis in Sec. \ref{Discussion_Conclusions}. 

\section{Setup}
In this section, we first provide an executive summary of white dwarf structures in beyond Horndeski theories of gravity and in scalar-tensor-vector gravity (or MOG theory). We then discuss the equation of state of the white dwarfs used in this study and provide a summary of the observational data.
\subsection{White dwarfs in beyond GR theories} 
\label{White_dwarfs_in_beyond_GR_theories}
The general class of beyond Horndeski theories of $G^3$ type are generally described by an action functional \cite{Zumalacarregui:2013pma, Gleyzes:2014dya, Gleyzes:2014qga, Kobayashi:2019hrl}
 \begin{equation}
S[g_{\mu\nu},\psi^{(i)}] = \int d^4 x\,\sqrt{-g}\mathcal{L}[g_{\mu\nu},\psi^{(i)}],   
 \end{equation}
in which the Lagrangian density $\mathcal{L}$ is a Lorentz-scalar which depends locally on the metric and matter fields ($g_{\mu\nu}$, $\psi^{(i)}$) and their derivatives. In these theories, the modified Tolman-Oppenheimer-Volkoff (TOV) equation for the hydro-static equilibrium of the star reads \cite{Koyama:2015oma, Barreira:2014jha, Sakstein:2015zoa, Saito:2015fza, Jain:2015edg}:
\begin{equation}\label{TOV1}
\frac{dP}{dr}=-\frac{Gm\rho}{r^2}-\frac{\gamma}{4}G \rho \frac{d^2m}{dr^2},
\end{equation}
where $P$ and $\rho$ are the pressure and energy density at the distance $r$ from the center of the star respectively and $m$ is the enclosed mass within the radius $r$. The dimensionless parameter $\gamma$ characterizes the effects of modification of gravity. The mass $m$ contained within the radius $r$ is calculated from the mass continuity equation:
\begin{equation}\label{mass_continuity1}
	\frac{dm}{dr}=4\pi r^2 \rho.
\end{equation}
We then rewrite Eq.\eqref{TOV1} using Eq.\eqref{mass_continuity1} as:
\begin{equation}\label{modified_TOV}
	\frac{dP}{dr}=-\frac{Gm\rho}{r^2}\left[1+\frac{\pi r^3 \gamma}{m}\left(2\rho +r\frac{d\rho}{dr}\right)\right].
\end{equation}
We note that, in Eq.\eqref{modified_TOV}, when $\gamma=0$, we recover the TOV equation in Newtonian gravity (i.e. in the weak-field limit of GR). To solve for the internal structure of a star, we then integrate Eq.\eqref{mass_continuity1} and Eq.\eqref{modified_TOV} simultaneously.
    
Scalar-tensor-vector gravity (STVG), otherwise known as modified gravitational (MOG) theory \cite{Moffat:2005si}, includes a massive vector field $\phi_{\mu}$ and three scalar fields $G$, $\mu$ and $\omega$. This theory has the following form of action \cite{Moffat:2005si, LopezArmengol:2016irf,Moffat:2020jic}: 

\begin{equation}\label{action1} 
S=\int d^4x \sqrt{-g}\left[\frac{1}{16\pi G}R+\frac{1}{4}B^{\mu\nu}B_{\mu\nu}\right]+S_{M},
\end{equation}
where $B^{\mu\nu}$ is the Faraday tensor of the vector field $\phi_{\mu}$ which is defined by $B_{\mu\nu}=\partial_{\mu}\phi_{\nu}- \partial_{\nu}\phi_{\mu}$, and $S_{M}$ is possible matter sources. The modified TOV equation for MOG theory reads \cite{Moffat:2005si, LopezArmengol:2016irf,Moffat:2020jic} :
\begin{align}\label{modified_TOV_equation_STVG}
\frac{dP}{dr}=&\frac{Q_{g}}{r^4}\frac{dQ_{g}}{dr} - \frac{\exp(\lambda)}{r^2} \left(\rho c^2+P\right)\nonumber\\ 
     & 	\times \left(\frac{4\pi G}{c^4} r^3P-\frac{2 G Q^2_{g}}{c^4 r} +\frac{G m}{c^2} +\frac{2\pi G}{c^4}\int dr \frac{Q^2_{g}}{r^2}\right)
\end{align}
where $G=G_{N}(1+\alpha)$, $Q_{g}=\sqrt{\alpha G_{N}} M$ and 
\begin{align}
&\exp(-\lambda(r)) =1-\frac{2Gm(r)}{c^2 r}-\frac{4\pi G}{r c^4}\int dr \frac{Q_{g}(r)}{r^2}.
\end{align}
The parameter $\alpha$ quantifies the effect of modification of gravity with $\alpha=0$ implying a pure GR scenario. Finally, solving for Eq.\eqref{mass_continuity1} and Eq.\eqref{modified_TOV_equation_STVG} gives us the structure of a star in MOG theory.

\subsection{Equation-of-state for white dwarfs}
\label{Equation_of_state}
In a white dwarf, the electron becomes degenerate. The resulting degeneracy pressure in the stellar interior balances the inward gravitational pull, and a hydrostatic equilibrium is achieved. Following the formalism developed in Ref. \cite{book1}, we use a simple model of carbon-oxygen white dwarfs at zero temperature. The model assumes that degenerate electron gas is in the ground state and the star is in a completely ionized state. For such white dwarfs, electron degeneracy pressure is given by \cite{book1}
\begin{align}\label{pressure}
P_{e}&=\frac{2}{3h^3}\int^{p_{F}}_{0}\frac{p^2c^2}{(p^2c^2+m^2_{e}c^4)^{1/2}}4\pi p^2dp \nonumber \\
&=\frac{8\pi m^{4}_{e}c^5}{3h^3}\int^{x}_{0}\frac{x^4 dx}{\sqrt{\left( 1+x^2 \right)}} \nonumber \\ 
&=1.4218\times10^{24}\phi(x) ~~{\rm N.m^{-1}}, 
\end{align}
where $p_{F}$ is the Fermi momentum, $x=\frac{p_{F}}{m_{e}c}$ is the dimensionless Fermi momentum, $m_{e}$ is the electron mass and
\begin{align}
\phi(x) = & \frac{1}{8\pi^2}\left\{  x\left(x^2 +1\right)^{1/2}\left(\frac{2 x^2}{3}-1\right) ~~~+  \right. \nonumber \\
&\left. \quad + \ln\left[ x+ \left(1+x^2\right)^{1/2}\right] \right\}.
\end{align}
On the other hand, the total energy density of the white dwarf is the sum of the energy density of electrons and non-relativistic carbon atoms, $\rho = \rho_{e} + \rho_{C}$. However, $\rho$ is completely dominated by $\rho_{C}$ that reads
\begin{align}\label{energy_density}
     \rho_{C} &= \frac{m_{C} n_{e}}{6}=1.9479\times 10^{9} x^3 ~~{\rm kg.m^{-3}},
\end{align}
where $m_{C}$ is the mass of ionized carbon and 6 is carbon's atomic number, and $n_{e}=\frac{x^3}{3\pi^2\lambda_{e}^3}$ is the number density of electrons with $\lambda$ being the electron Compton wavelength.
Since the pressure from carbon $P_{C} \ll P_{e}$, the total pressure becomes $P \simeq P_{e}$. 
Therefore, once we know $x$, we can compute both the energy density $\rho_C$ (using Eq.\eqref{energy_density}) and the pressure $P_e$ (using \eqref{pressure}). In practice, we know either the density or the pressure of the white dwarf. We then calculate $x$ either from Eq.\eqref{energy_density} or Eq.\eqref{pressure}. Finally, from $x$, we obtain the other quantity.

\subsection{Description of the observational data}
\label{sec:data}
The mass and radius data of several white dwarfs, along with their respective errors, specially in binary systems, have been determined through various astrophysical probes. We use the mass-radius data for 63 white dwarfs from \textit{Gaia} DR1 \cite{10.1093/mnras/stw2854} and \textit{Hipparcos} \cite{10.1093/mnras/stw2854}. Samples are obtained by combining measurements of white dwarf parallaxes and spectroscopic atmospheric parameters. We consider white dwarfs that have been observed directly and those observed in wide binaries. Furthermore, we extract the white dwarf's mass-radius data from Ref. \cite{Holberg:2012pu} (Holberg \textit{et al}), which are derived from spectroscopic temperatures and gravity, and Ref. \cite{10.1093/mnras/stx1522} (Parsons \textit{et al}), which are derived from the
photometric observations of the eclipses and kinematic parameters, and are almost completely independent of white dwarf model of the atmospheres.

\section{Results}
\label{Result}
In this section, we now briefly summarise white dwarf structures in modified theories of gravity and test theoretically predicted white dwarf mass-radius relation against the observational data.
\subsection{White dwarf structures}
We first investigate the internal structure of white-dwarfs by solving the modified TOV equation (Eq.\eqref{modified_TOV} for beyond Horndeski class of theories and Eq.\eqref{modified_TOV_equation_STVG} for MOG) and the mass continuity equation (Eq.\eqref{mass_continuity1}) simultaneously for a given theory of gravity. We assume the white dwarfs to be at zero temperature and in hydro-static equilibrium. The details of the chosen equation of state is given in Sec.\ref{Equation_of_state}.

We take the initial conditions to be $m(r=0)=0$ and $\rho(r=0)=\rho_c$ where $\rho_c$ is the central density. The pressure at the center of the star is calculated from the central density using the equation of state. We further assume that the density of the white dwarf is the same within a small radius $r_{0}$. This allows us to write pressure and mass at a distance $r_0$ from the center as: $P{(r_{0})}=P(\rho_{c})$ and $m{(r_{0})}=\frac{4\pi}{3}\rho_{c}r_{0}^{3}$.
We then solve the modified TOV equations (and the mass continuity equation) inside out until the pressure $P(r)$ becomes zero. The point where $P(r)$ becomes zero determines the radius $R$ of the white dwarf.
The mass contained within the radius $R$ gives the total mass of the star and is denoted by $M$.
This process gives us the mass, density and pressure profiles of the white dwarfs as a function of the radius.

\begin{figure}
\includegraphics[width=\columnwidth]{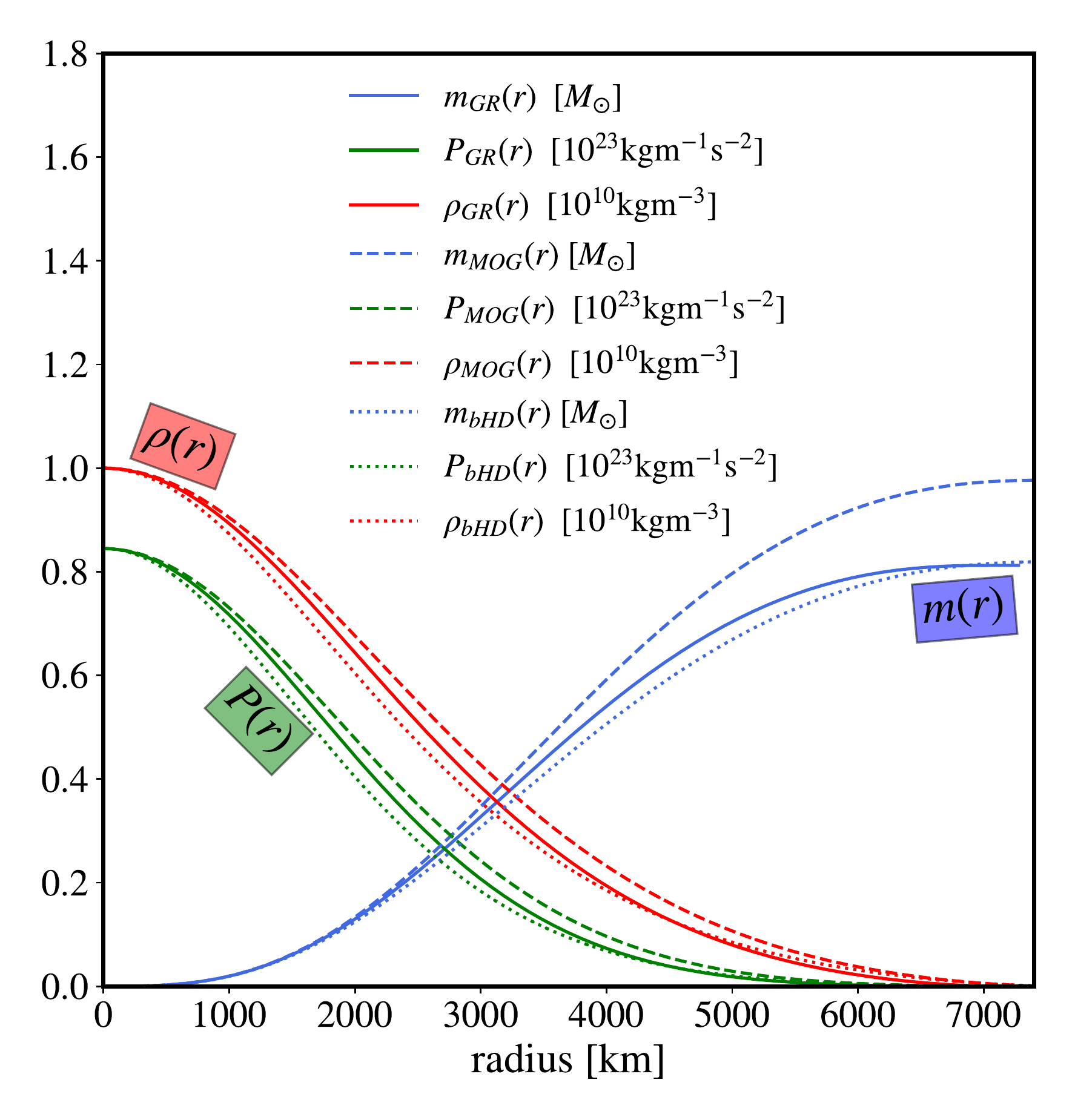}
\caption{We show the mass $m(r)$ (blue), pressure $P(r)$ (green) and density $\rho(r)$ (red) at a distance $r$ from the center of the white dwarf for a particular central density $\rho_{c}=10^{10} kg.m^{-3}$ in three different theories of gravity: GR (solid lines; $\alpha=0.0$), MOG theory (dashed lines; $\alpha=0.01$) and beyond Horndeski theory of type $G^{3}$ (dotted lines; $\gamma=0.15$; referred as \textit{bHD}). The details are in the text.
}
\label{fig:example_WD}
\end{figure}

Figure \ref{fig:example_WD} shows the mass profile $m(r)$ (blue lines), density profile $\rho(r)$ (red lines) and the pressure profiles $P(r)$ (green lines) for a representative white dwarf in both MOG theory (dashed lines) as well as in beyond Horndeski theory of type $G^3$ (dotted lines).
We choose the central density to be $\rho_{c}=10^{10} kg.m^{-3}$. As expected, the mass $m(r)$ increases as the distance from the center, $r$, increases while both the pressure $P(r)$ and the density $\rho(r)$ decrease with $r$.
For MOG theory, we consider $\alpha$, the parameter that controls the strength the modification of gravity, to be 0.01. For beyond Horndeski theory of type $G^3$, we choose $\gamma$, the parameter that controls the strength the modification of gravity, to be 0.15.
For comparison, we also show the corresponding GR predictions (obtained by setting $\alpha=0$ in MOG theory) in solid lines.
We find that the white dwarf mass is $M=0.819~M_\odot$ in beyond Horndeski theories, $M=0.976~M_\odot$ in MOG and $M=0.81~M_\odot$ in GR while the radius is $R=7836$ km in beyond Horndeski theories, $R=7737$ km in MOG and $R=7276$ km in GR.

\begin{figure}
\includegraphics[width=\columnwidth]{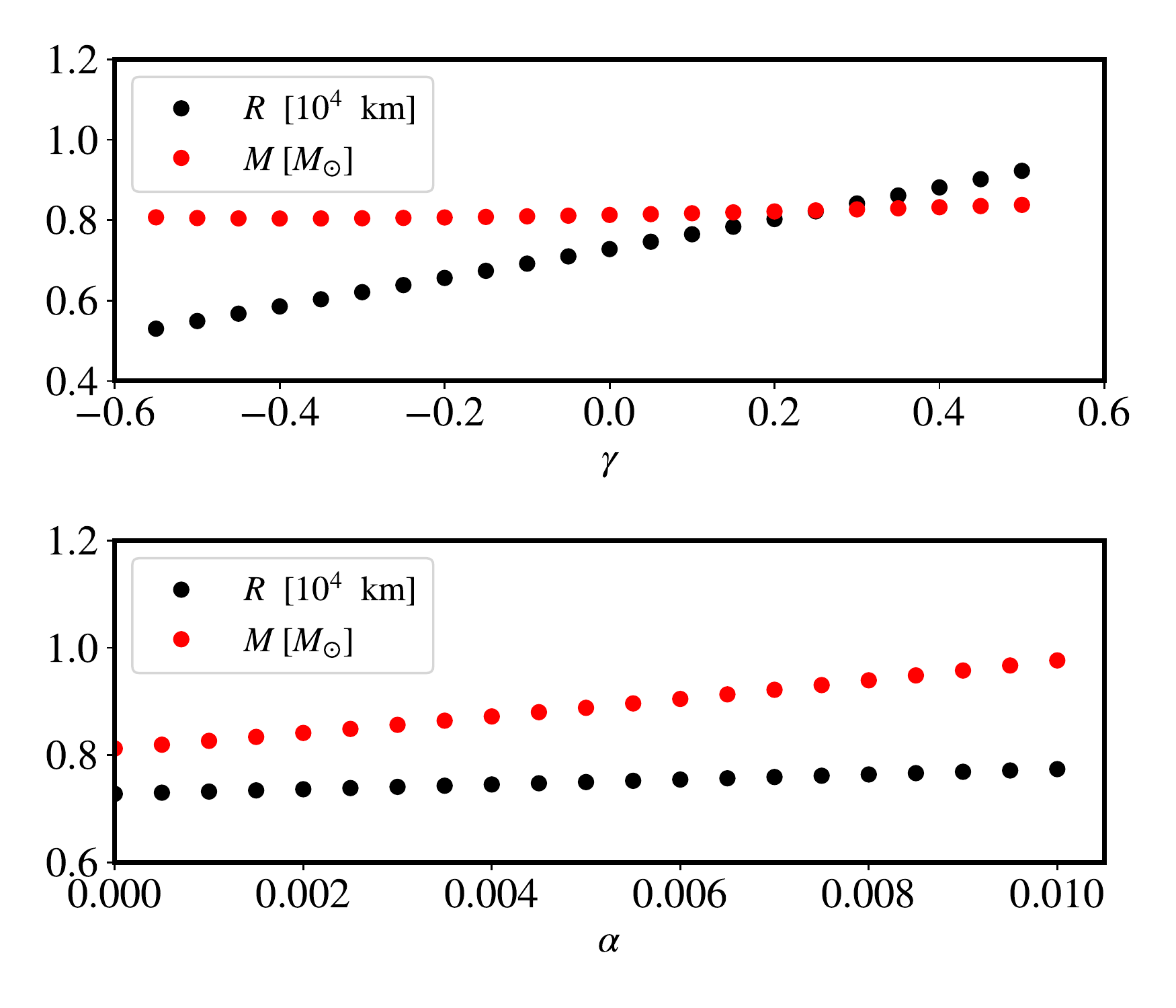}
\caption{We show the mass $M$ and the radius $R$ of white dwarfs as a function of the modification parameter $\gamma$ ($\alpha$) for the beyond Horndeski theories of $G_3$ type (scalar-tensor-vector gravity) for a fixed central density $\rho_{c}=10^{10} kg.m^{-3}$ in the upper panel (lower panel).}
\label{fig:MR_vs_gamma_alpha}
\end{figure}

We conclude this section by showing how the masses and the radii of the white dwarf change as a function of the deviation parameters $\gamma$ and $\alpha$ in beyond Horndeski theories of $G_3$ type and in scalar-tensor-vector gravity respectively (Fig.~\ref{fig:MR_vs_gamma_alpha}) for a fixed value of the central density $\rho_{c}=10^{10} kg.m^{-3}$. We observe that as $\gamma$ increases, white dwarfs become slightly more massive and its radius increases almost linearly. For MOG, we find that the dependence of the radii on the deviation parameter $\alpha$ is weak while the mass of the white dwarf changes quite linearly with $\alpha$.

\begin{figure}
\includegraphics[width=\columnwidth]{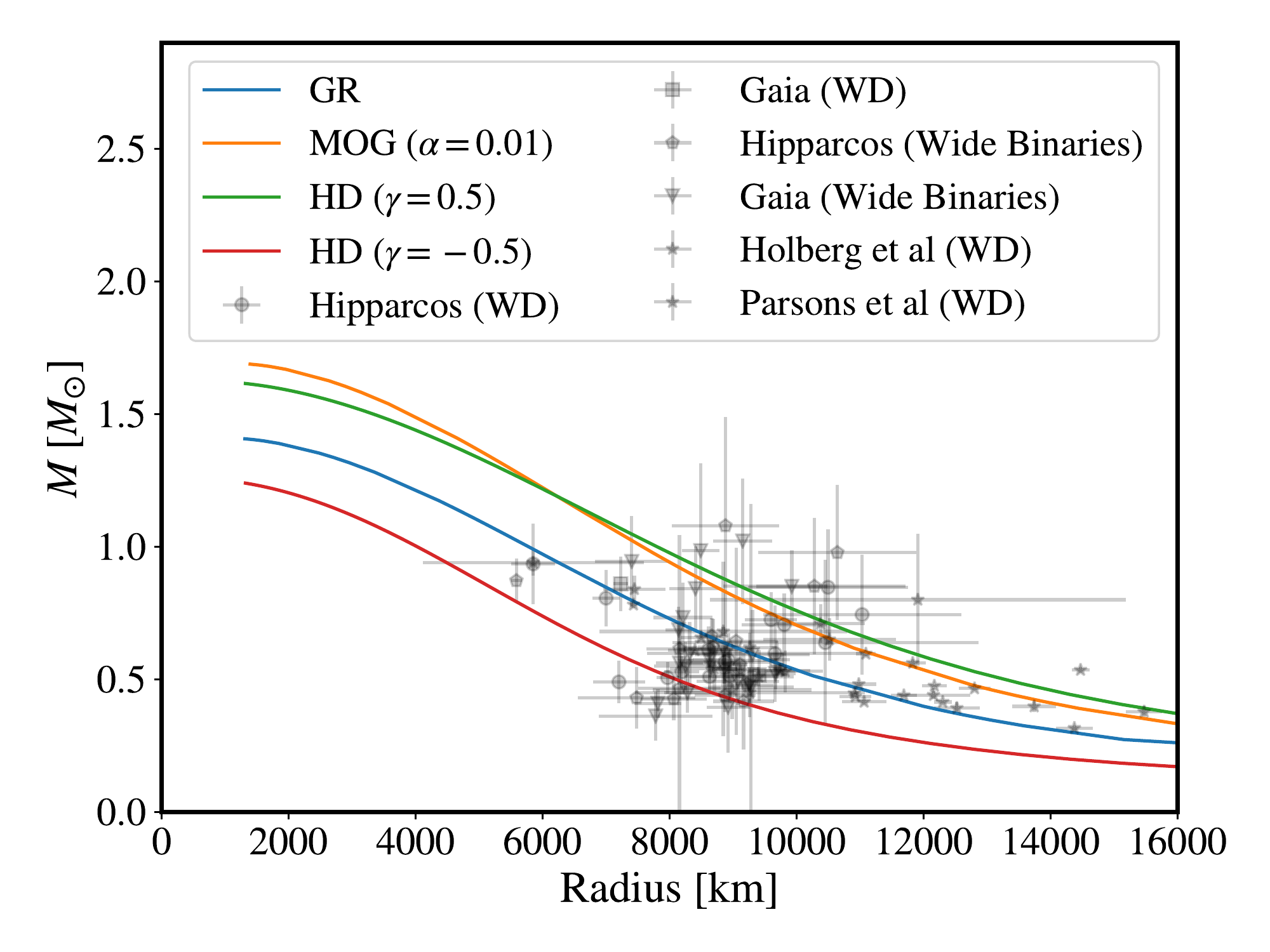}
\caption{We show the theoretical mass-radius curves for the white dwarfs in different theories of gravity (GR in blue, MOG in orange and beyond Horndeski theory of type $G^{3}$ in green/red respectively) and the astrophysical mass-radius data compiled in Section \ref{sec:data} (details are in text).}
\label{fig:MR_Data_Prediction_MOG_HD}
\end{figure}

\subsection{Testing mass-radius relation}
We, now, validate our numerical results against the mass-radius data compiled in Section \ref{sec:data} from five different astronomical surveys. 
In Fig \ref{fig:MR_Data_Prediction_MOG_HD}, we show the data with associated error bar as well as the theoretically predicted white dwarf mass-radius curves for both the beyond Horndeski theory of type $G^{3}$ and MOG theory.
To compute the theoretical mass-radius relation curves, we consider a range of values for the central density $\rho_c$ of the white dwarf from $3\times10^{8} kg.m^{-3}$ to $10^{13} kg.m^{-3}$, typical for astrophysical white-dwarfs. For each value of $\rho_c$, we obtain the total mass $M(R)$ and the radius $R$ of the white dwarf following the methods mentioned in the previous section. We then build cubic spline of $M$ as a function of $R$ using \texttt{scipy.interpolte.splrep}~\cite{splrep:abcd} and \texttt{scipy.interpolte.splrev}~\cite{splrep:efgh} to compute the corresponding white dwarf mass for any arbitrary value of $R$.

For the MOG theory, we show the predicted mass-radius relation for a fiducial value of $\alpha=0.01$ (orange line). 
For beyond Horndeski class of theories, we choose two different values of the modification parameters: $\gamma=0.5$ (green line) and $\gamma=-0.5$ (red line). 
Finally, we include the mass-radius relation curve computed within GR (blue line) for comparison.
We find that the qualitative behaviour of the mass-radius relation is the same across different theories of gravity. 
However, the total mass of the white dwarfs are quite different depending on the theory of gravity and the value of the modification parameters ($\alpha$ and $\gamma$).
In beyond Horndeski class of theories, negative values of $\gamma$ yield less massive white dwarfs than in GR and positive values of $\gamma$ predicts more massive white dwarfs.

\subsubsection{$\chi^2$ statistics}
To understand how well the astrophysical data matches the theoretical predictions, we compute the $\chi^2$  values between them. For a set of astrophysical mass-radius data-set $\{R_{i},M{_i}(R_{i})\}$ and the corresponding theoretical predictions $\{R_{i},M_{th}(R_{i})\}$, $\chi^2$ value is defined as
\begin{equation}
\chi^2=\sum_{i}^{N}\Delta\chi^2_{i}(R_{i}), \label{chi_square_equation}
\end{equation}
where 
\begin{equation}
\Delta \chi^2_{i}(R_{i}) = \frac{(M_{th}(R_{i})-M_{i})^2}{\sigma^2_{M,i}}
\end{equation}
and $N$ is the total number of observational white dwarfs. The d.o.f. is the number of degrees of freedom, which is $(2N-n-1)$. Here, the factor $2$ comes from the fact that we have two independent observations for each white dwarf, and n is the number of fitting parameters. The value of $n$ is unity for both the beyond Horndeski theory of type $G^{3}$ and MOG theory. Smaller values of $\chi^2$ indicate a better match between data and prediction while larger values signify deviations from the observed data. The best fit values of the modification parameters $\alpha$ or $\gamma$ corresponds to the minima in the $\chi^2$ curve. 

\subsubsection{Best-fit values for $\alpha$ and $\gamma$}
\begin{figure}
\includegraphics[width=\columnwidth]{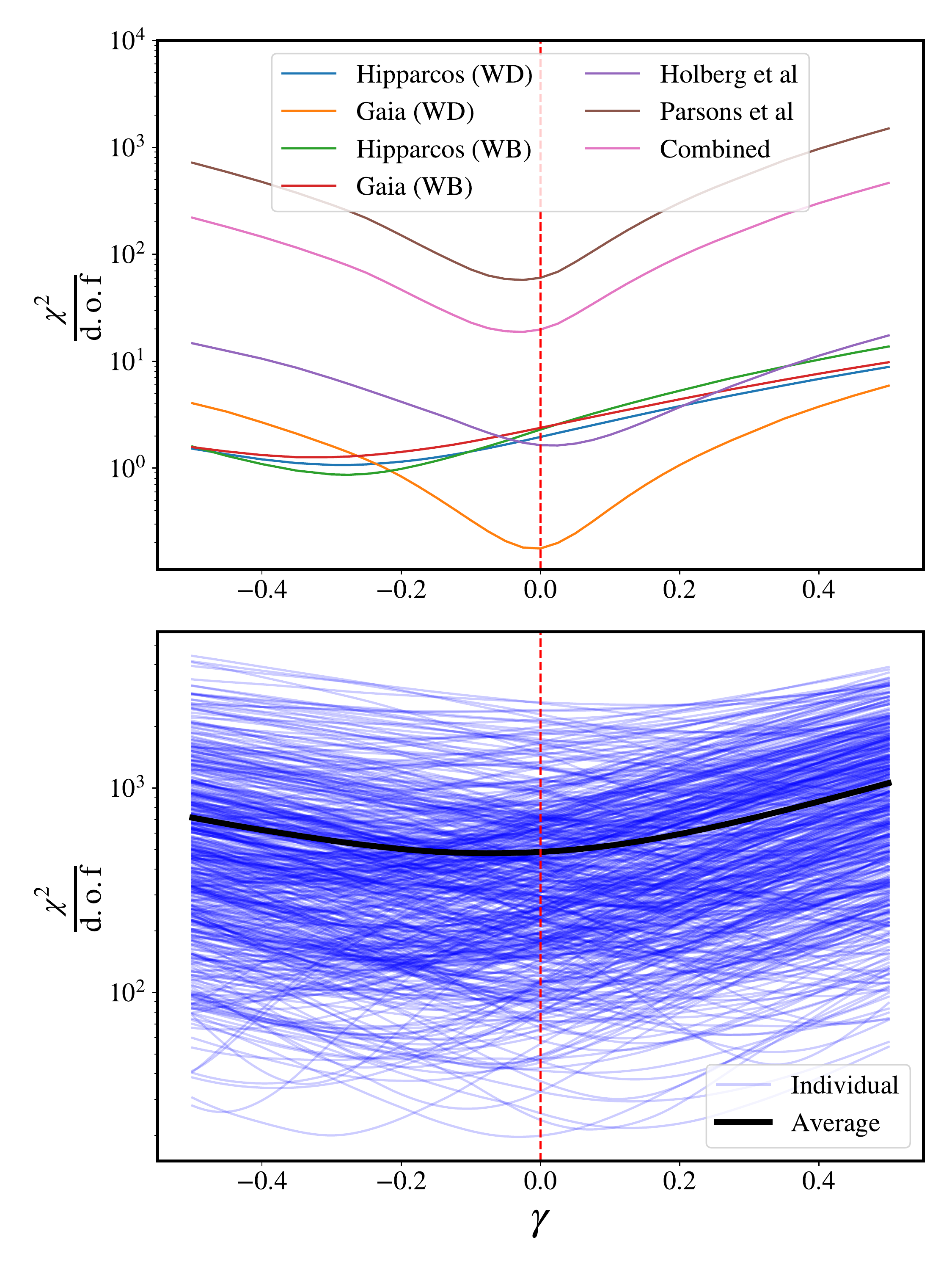}
\caption{\textit{Upper panel :} $\chi^2$ curves for the beyond Horndeski theory of type $G^{3}$ computed using six different astrophysical data-set as well as from the combined data-set. 
\textit{Lower panel :} a selection of 100 $\chi^2$ curves (blue) for the beyond Horndeski theory of type $G^{3}$ computed using a data-set containing randomly selected 25 points from the combined data-set.
Solid black line is the averaged $\chi^2$ curve obtained from the 100 $\chi^2$ curves shown in blue.
Red vertical line denotes a $\chi^2$ value of zero obtained when the data matches the theoretical predictions exactly. The details are in the text.}
\label{fig:chisq_curves_HD}
\end{figure}

To find out the best-fit value of $\gamma$ for the beyond Horndeski theory of type $G^{3}$, we perform the $\chi^2$ fitting using the combined set of data containing 63 points and show the results in Fig. \ref{fig:chisq_curves_HD} (upper panel). We find that the minimum in $\chi^2$ occurs at $\gamma=-0.025$, close to the Newtonian gravity expection of $\gamma=0$. At this point, we ask the question whether the best-fit $\gamma$ value can change due the selection effect in the astrophysical data. To understand that, we repeat our analysis for all individual data-sets. We observe that, for three of the individual datasets (white dwarf data obtained from the wide binary systems in the Hipparcus survey, from the wide binary systems in Gaia survey and from observing the individual white dwarfs directly in the Hipparcus survey respectively), the best-fit value of $\gamma$ deviates from zero.

To further understand the selection effect, we perform the following experiment. We select a total of 25 mass-radius data points from the combined set of 63 points and perform the $\chi^2$ fit and obtain the best-fit value. We then repeat this step 100 times to emulate 100 independent astrophysical survey data. Fig. \ref{fig:chisq_curves_HD} (lower panel) shows all 100 different $\chi^2$ curves as a function of $\gamma$ as well as the averaged $\chi^2$ curve (bold black line). We call the averaged $\chi^2$ curve as the bootstrap average. The plot features cases where the minima of the $\chi^2$ curves are far away from $\gamma=0$ with no strong preference for the positive or negative values of $\gamma$. The best-fit value of $\gamma(=-0.075)$ obtained from the averaged $\chi^2$ curve is close to zero.

\begin{figure}
\includegraphics[width=\columnwidth]{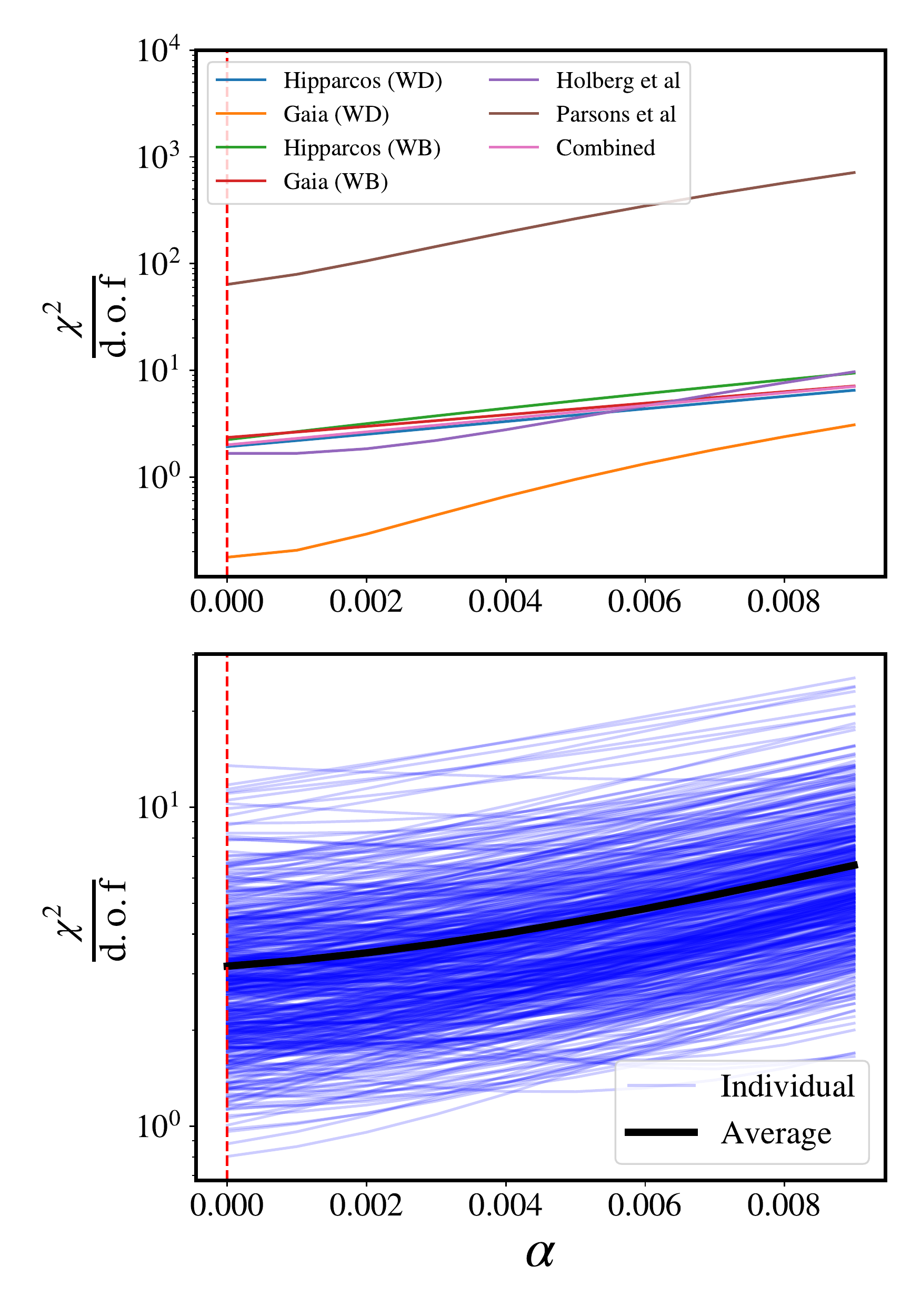}
\caption{\textit{Upper panel :} $\chi^2$ curves for the MOG theory computed using six different astrophysical data-set as well as from the combined data-set. 
\textit{Lower panel :} a selection of 100 $\chi^2$ curves (blue) for the MOG theory computed using a data-set containing randomly selected 25 points from the combined data-set.
Solid black line is the averaged $\chi^2$ curve obtained from the 100 $\chi^2$ curves shown in blue.
Red vertical line denotes a $\chi^2$ value of zero obtained when the data matches the theoretical predictions exactly. 
The details are in the text.}
\label{fig:chisq_curves_MOG}
\end{figure}

We then repeat the analysis for the MOG theory and show the resultant $\chi^2$ curves as a function of the modification parameter $\alpha$ in Fig. \ref{fig:chisq_curves_MOG}. Unlike $\gamma$ in beyond  Horndeski theory of type $G^{3}$, $\alpha$ can not be negative. We find that the overall qualitative results are similar to the ones obtained for the beyond Horndeski theory of type $G^{3}$. Our result suggests that the selection effect in data may be important to understand while interpreting the best-fit parameters in modified gravity theories from white-dwarf mass-radius relation.

\subsubsection{90\% credible intervals}
\begin{figure}
\includegraphics[width=\columnwidth]{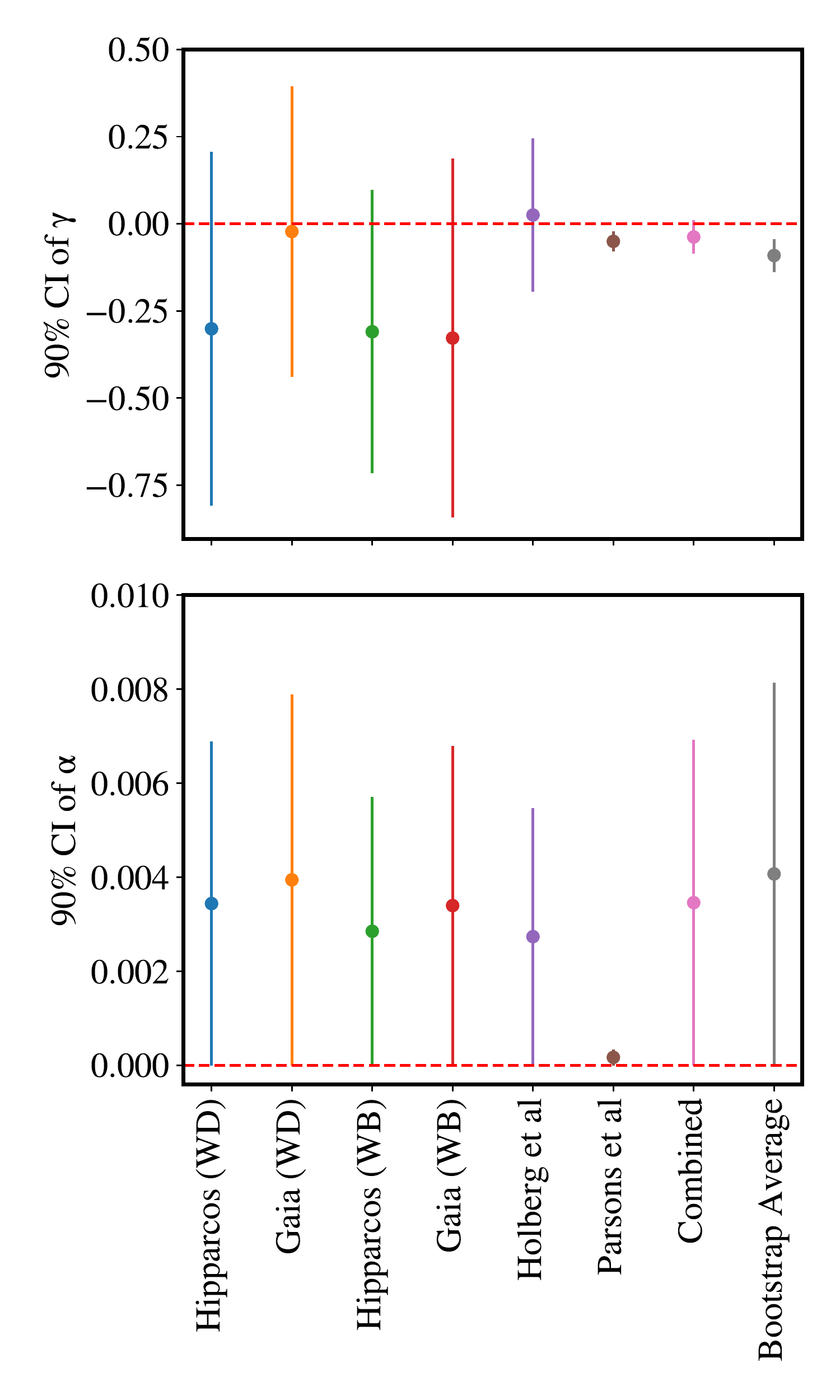}
\caption{We show the constraints on the deviation parameters $\gamma$ ($\alpha$) for the beyond Horndeski theory of type $G^{3}$ (MOG-theory) at the 90\% credible intervals obtained from different sets of astrophysical data.}
\label{fig:90CI}
\end{figure}

Finally, we compute the likelihood of the modification parameters $\alpha$ and $\gamma$ from their respective $\chi^2$ values as:
\begin{equation}
    \mathcal{L}(x) \propto e^{-{(\chi^2/2)}}.
    \label{eq:likelihood}
\end{equation}
The likelihood function $\mathcal{L}(x)$ gives the probability of a particular value of $x$ (in our case,  $\alpha$ or $\gamma$) describing the data. The most probable value of $x$ maximizes the likelihood function. This condition is, therefore, equivalent to minimizing the $\chi^2$ values. However, the advantage of using the likelihood function is that it will allow us to compute the best-fit values of $x$ (in our case,  $\alpha$ or $\gamma$) with its associated 90\% credible intervals. In Fig. \ref{fig:90CI} (upper panel), we show the best-fit $\gamma$, for the beyond Horndeski theory of type $G^{3}$, along with the error bars at the 90\% credible intervals obtained from using different sets of astrophysical data. We find that the data-set from Parsons \textit{et al}, the combined data-set and the bootstrap average yields the most constrained measurement of $\gamma$. As a sanity check, we further confirm that our measurement of $\gamma$ from the data-set compiled in Parsons \textit{et al} matches to the values obtained by Jain \textit{et al}~\cite{Jain:2015edg}. We then repeat the analysis for the MOG theory and provide the best-fit $\alpha$ values using different dataset in Fig. \ref{fig:90CI} (lower panel).

\section{Discussion \& Conclusions}
\label{Discussion_Conclusions}
In this work, we have explored the imprints of the modification of gravity on the stellar structure of the white dwarfs. In particular, we have studied two distinct classes of the alternative theories of gravity: (i) scalar-tensor-vector gravity and (ii) beyond Horndeski theories of $G_3$ type. Both the theories have been studied extensively as viable alternatives to general relativity in explaining astrophysical observations at the galactic and extra-galactic scales. We observe that the modification of the underlying theory of gravity changes some of the white dwarf observable such as the masses and the radii. We then use the observed mass-radius relation data from the white dwarfs to test the predicted mass-radius relation in both the theories. As a part of our analysis, we also provide updated constraints on $\gamma$ and $\alpha$, parameters controlling the deviation of the gravitational force from the GR values in beyond Horndeski theories of $G_3$ type and  scalar-tensor-vector gravity theories respectively. Our best-fitted values are: $0.007 \le \alpha$ for the scalar-tensor-vector gravity and $-0.08 \le \gamma \le 0.007$ for the beyond Horndeski theories of $G_3$ type.

One of our interesting observations is that the best fitted values for $\gamma$ and $\alpha$ change significantly depending on the astrophysical data used. This motivates us to dig into the selection effect in data from different astrophysical catalogs on the test of gravity using white dwarfs. 
To do this, we randomly select a subset of the all astrophysical data to imitate different catalogs of white dwarf mass radius relation data and obtain the best-fitted $\gamma$ and $\alpha$ respectively. We repeat this exercise multiple times to obtain the average best-fitted $\gamma$ and $\alpha$. We find that the average best-fitted $\gamma$ and $\alpha$ obtained this way is very close to the GR expectation while $\gamma$ and $\alpha$ values obtained from some of the individual data-set may exhibit sign of a deviation from GR. Our results therefore ask for caution while interpreting results of any exercise involving the test of gravity using astrophysical data (specially for the white dwarfs). We stress that it is important to understand the systematic of the data collection in order to combine mass-radius relation data from different astrophysical probes. While this is important for developing more stringent tests of gravity using white-dwarfs, it will require further investigation which is beyond the scope of this work.

At this point, it is important to note that our treatment of the white dwarfs is very simplistic in nature. We have assumed simple carbon-oxygen white dwarfs structure which is not always true. In particular, we know of some low mass Helium white dwarfs which are created by Hydrogen burning through carbon-nitrogen-oxygen cycle \cite{Istrate:2016grc}. Furthermore, we assume a zero temperature equation of state for the white dwarfs. In reality, white dwarfs may have finite temperature equation of state \cite{DKoester1990}. In future, it will be interesting to understand whether assuming finite temperature equation-of-states may change any of our results (and results appeared elsewhere) significantly. We leave that as an important future work.

While this paper focuses on the white dwarfs, one can extend this to other compact objects such as the brown dwarfs and the neutron starts~\cite{Hossain:2021qyg,Sakstein:2015zoa}. Particularly, it will be interesting to cross-correlate the best-fitted deviation parameters obtained using neutron star data in the electromagnetic window and the ones obtained from gravitational waves observations. 


\begin{acknowledgments}
The authors would like to thank Prof. Koushik Dutta for insightful comments and discussions on this project. 
T.I. is supported by NSF Grants No. PHY-1806665 and DMS-1912716. Part of this work is additionally supported by the Heising-Simons Foundation, the Simons Foundation, and NSF Grants Nos. PHY-1748958.

\end{acknowledgments}  

\bibliography{References.bib}

\end{document}